\documentclass[aps,prl,floats,twocolumn,showpacs,superscriptaddress]{revtex4}

\usepackage{graphicx,epsfig}
\usepackage{rotate}

\usepackage{times}
\usepackage{graphics,dcolumn,bm,fleqn,epic,eepic,float}
\usepackage{amssymb,amsmath,multirow,rotate,color}

\begin{document}

\title{Synchronization of networks with variable local properties}

\author{Jes{\'u}s G{\'o}mez-Garde\~{n}es}

\affiliation{Institute for Biocomputation and Physics of Complex
Systems (BIFI), University of Zaragoza, Zaragoza 50009, Spain}

\affiliation{Departamento de F\'{\i}sica de la Materia Condensada,
University of Zaragoza, Zaragoza E-50009, Spain}

\author{Yamir Moreno}

\affiliation{Institute for Biocomputation and Physics of Complex
Systems (BIFI), University of Zaragoza, Zaragoza 50009, Spain}

\date{\today}

\begin{abstract}

We study the synchronization transition of Kuramoto oscillators in
scale-free networks that are characterized by tunable local
properties. Specifically, we perform a detailed
finite size scaling analysis and inspect how the critical properties
of the dynamics change when the clustering coefficient and the average
shortest path length are varied. The results show that the onset of
synchronization does depend on these properties, though the dependence
is smooth. On the contrary, the appearance of complete synchronization is
radically affected by the structure of the networks. Our study
highlights the need of exploring the whole phase diagram and not only
the stability of the fully synchronized state, where most studies have
been done up to now.

\end{abstract}


\maketitle

\section{Introduction}

Emergent collective phenomena have been studied since long time
ago. These phenomena arise in many fields of science, ranging from
natural to social and artificial systems. They are characterized,
among other features, by the collective behavior of many interacting
units that show a pattern hard to predict from the individual
behavior of the system constituents. Several seminal models of
statistical physics and non-linear dynamics have been scrutinized as
paradigms of self-organization, emergence and cooperation between the
units forming the system. In particular, synchronization phenomena
constitute one of the most striking examples because of the many
systems showing synchronization patterns in their behavior
\cite{winfree,strogatzsync,zanette}.

One of the most celebrated synchronization models is due to Kuramoto
\cite{kurabook,conradrev}, who analyzed a model of phase oscillators
coupled through a function (sine) of their phase differences. This
model owes most of its success to the plenty of analytical insights
that one can get through the mean-field approximation originally
proposed by Kuramoto. In this approach (KM), the nodes of an all to all,
i.e. globally, coupled network, are considered to be oscillators with
an intrinsic frequency and their phases evolve in time in such a way
that if the coupling between them is larger than a critical threshold,
the whole system gets locked in phase and attains complete
synchronization.

However, it has been recently discovered that real systems do not show
a homogeneous pattern of interconnections among their parts. That is,
the underlying structure is not compatible with the original
assumption of the KM. It is not even well described by random patterns
of interconnections in the vast majority of systems. Therefore, the
mean-field approach requires of several constraints that are not
usually fulfilled in real systems. Natural, social and technological
systems show complex patterns of connectivity that characterize
seemingly diverse social \cite{pnas}, biological \cite{ref5,ref6} and
technological systems \cite{ref7,wang}. They exhibit common features
that can be captured using the tools of graph theory or in more recent
terms, network modeling \cite{book1,book2,yamirrep}. 

It turns out that many real networks are well described by the
so-called scale-free (SF) networks. Their main feature is that the
probability that a given node has $k$ connections to other nodes
follows a power-law $P_k\sim k^{-\gamma}$, with $2 \le \gamma \le 3$
in most cases \cite{book1,yamirrep}. The study of processes taking
place on top of these networks has led to reconsider classical results
obtained for regular lattices or random graphs due to the radical
changes of the system's dynamics when the heterogeneity of complex
networks can not be neglected
\cite{book1,book2,yamirrep,pv00,pastor-satorras-2001,mpv02,ceah00}.

It is then natural to investigate how synchronization phenomena in
real systems are affected by the complex topological patterns of
interaction. This is not an easy task, as one has to deal with two
sources of complexity, the nonlinear character of the dynamics and
highly non trivial complex structures. In recent years, scientists
have addressed the problem of synchronization capitalizing on the
Master Stability Function (MSF) formalism \cite{pecora} which allows
to study the stability of the {\em fully synchronized state}
\cite{barahona,motter1,motter2,munozprl,zhou,prl_stefano}. While the MSF approach
is useful to get a first insight into what is going on in the system
as far as the stability of the synchronized state is concerned, it
tells nothing about how synchronization is attained and whether or not
the system under study exhibits a critical point similar to the
original KM. To this end, one must rely on numerical calculations and
explore the {\em entire phase diagram}. Surprisingly, there are only a
few works that have dealt with the study of the whole synchronization
dynamics in specific scenarios
\cite{yamir,kahng,mcgraw,arenas,jma01,jma02} as compared with those
where the MSF is used, given that the onset of synchronization is
reacher in its behavioral repertoire than the state of complete
synchronization.

In this paper, we take a further step in the detailed characterization
of the phase diagram and specifically, in the description of the
dynamical behavior at the onset of synchronization in SF networks. By
performing a standard finite size scaling analysis, we show that the
local topology affects the critical properties of the dynamics, though
it is less pronounced than what one may expect a priori. We capitalize
on a network model that keeps the power-law exponent fixed while
varying the clustering coefficient and the average path length. In
what follows, we describe the topological and dynamical model and
discuss the results from a global perspective. Finally, in the last
section, we state our conclusions.

\section{Network model and Dynamics}

We implement a network model in which the graph is grown at each time
step by linking preferentially new nodes to already existing nodes in
the same way as in the Barabasi and Albert (BA) model
\cite{bar99}. The only difference is that the nodes are assumed to
have a fitness that characterizes their affinities \cite{gm}. In this
way, by tuning a single parameter $\mu$, one can go from the BA limit
down to a network in which several network properties vary as a
function of $\mu$. On the other hand, as the linking mechanism is
still the BA preferential attachment rule, the exponent $\gamma$ of
the power-law degree distribution is the same (i.e., $\gamma=3$)
regardless of the value of $\mu$. Roughly speaking, the model mimics
the situation in which new nodes are attached to an existing core or
network but without having knowledge of the whole topology.

The recipe is then as follows \cite{gm}. {\it i)} Initially, there is
a small, fully connected, core of $m_0$ nodes. Assign to each of these
$m_0$ nodes a random affinity $a_i$ taken from a probability
distribution. In this work, we have used for simplicity a uniform
distribution between $(0,1)$. {\it ii)} At each time step, a new node
$j$ with a random affinity $a_j$ is introduced and $m$ links are
established with nodes already present in the network following
the rule
\begin{equation}
\Pi(k_i)=\frac{k_i}{\sum_{s\in{\Gamma}}k_s},
\label{eq1}
\end{equation}
where the set $\Gamma$ contains all nodes that verify the condition
$a_i-\mu \le a_j \le a_i+\mu$, being $\mu \in(0,1)$ a parameter that
controls the affinity tolerance of the nodes. Finally, {\it iii)}
repeat step ({\it ii}) $t$ times such that the final size of the network
be $N=m_0+t$.

\begin{figure}[t]
\begin{center}
\epsfig{file=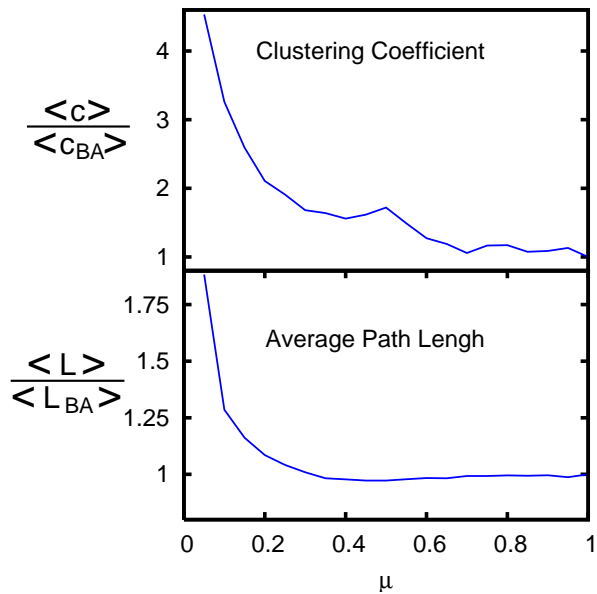,width=5.0in}
\end{center} 
\caption{(color online) Top: Values of the clustering coefficient
  relatives to those of the BA model against the parameter
  $\mu$. Bottom: The average cluster length for the network generated
  relatives to the BA value as a function of $\mu$. All the networks
  are made up of $N=1000$ and have an average degree $\langle k
  \rangle=6$.}
\label{fig1}
\end{figure}

In the above model, when $\mu$ is close enough to 1, the BA model is
recovered. When it is decreased from $1$, the values of some
magnitudes such as the clustering coefficient ($\langle c \rangle$)
and the average path length ($\langle L \rangle$) grows with respect
to the BA limit \cite{gm}. In Fig.\ \ref{fig1} we have represented how
these properties vary as a function of the parameter $\mu$. Note that
the larger variations correspond to the clustering coefficient (a
factor greater than 4 as compared to a factor close to 2 for $\langle
L \rangle$) and that it is the first property that deviates from the
BA limit. This tendency holds up to very small values of $\mu$, where
$\langle L \rangle$ raises at a higher rate than $\langle c \rangle$
(not shown in Fig.\ \ref{fig1}). More important for our purposes is
the region of $0.4 \le \mu \le 1$. For these values of $\mu$ one
observes that $\langle L \rangle$ remains constant while $\langle c
\rangle$ starts to grow as soon as it moves away from the BA
limit($\mu=1$). This allows to decouple the effects of both magnitudes
on what we are going to study. As we shall latter see, the structural
clustering plays a major role in the synchronization of Kuramoto
oscillators, as does in other dynamical processes \cite{echenique}.

The dynamic ingredient of the model is given by the collective
behavior that arises when the nodes are considered to be phase
oscillators that follow the Kuramoto model. In this formalism, the
population of $N$ interconnected units are coupled phase oscillators
where the phase of the $i$-th unit, denoted by $\theta_i(t)$, evolves
in time according to
\begin{equation}
\frac{d\theta_i}{dt}=\omega_i + \sum_{j}
\Lambda_{ij}A_{ij}\sin(\theta_j-\theta_i) \hspace{0.5cm} i=1,...,N
 \label{ks}
\end{equation}
\noindent where $\omega_i$ stands for its natural frequency,
$\Lambda_{ij}=\lambda$ \cite{note1} is the coupling strength between
units and $A_{ij}$ is the connectivity matrix ($A_{ij}=1$ if $i$ is
linked to $j$ and $0$ otherwise). Note that in the original Kuramoto
model mean-field interactions were assumed which leads to
$\Lambda_{ij}={\cal K}/N \forall i,j$, for the all-to-all
architecture. On the other hand, the model can be solved in terms of
an order parameter $r$ that measures the extent of synchronization in
a system of $N$ oscillators as:
\begin{equation}
re^{i\Psi}=\frac{1}{N}\sum_{j=1}^{N} e^{i\theta_j}
 \label{r_kura}
\end{equation}
\noindent where $\Psi$ represents an average phase of the system. The
parameter $r$ takes values $0\le r \le 1$, being $r=0$ the value of
the incoherent solution and $r=1$ the value for total synchronization.

\section{Results}

In order to inspect how the dynamics of the $N$ oscillators depends on
the underlying topology, we have performed extensive numerical
simulations of the model. Starting from $\lambda=0$, we increase at
small intervals its value. The natural frequencies and the initial
values of $\theta_i$ are randomly drawn from a uniform distribution in
the interval $(-1/2,1/2)$ and $(-\pi,\pi)$, respectively. Then, we
integrate the equations of motion Eq.\ (\ref{ks}) using a $4^{th}$
order Runge-Kutta method over a sufficiently large period of time to
ensure that the system reaches the stationary state, where the order
parameter $r$ is computed. The procedure is repeated gradually
increasing $\lambda$.

\begin{figure}[t]
\begin{center}
\epsfig{file=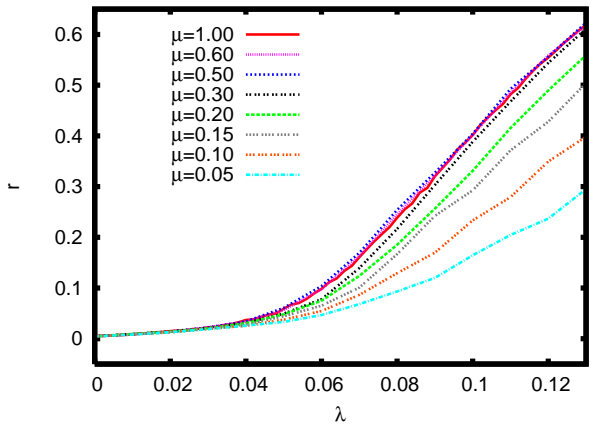,width=2.3in,angle=-90}
\end{center} 
\caption{(color online) Order parameter $r$ as a function of $\lambda$
for different values of $\mu$ as indicated. The network parameters are
those of Fig.\ \ref{fig1}.}
\label{fig2}
\end{figure}



The results for $r$ are shown in Fig.\ \ref{fig2} against the control
parameter $\lambda$ for several networks characterized by different
$\mu$. For all values of $\mu$, when the coupling is increased from
small values, the incoherent solution prevails and macroscopic
synchronization is not attained. This behavior persists until a
certain critical value $\lambda_c(\mu)$ is crossed. At this point some
elements lock their relative phase and synchronized nodes form. This
constitutes the onset of synchronization. Beyond this value, the
population of oscillators splits into a partially synchronized state
contributing to $r$ and a group of nodes whose natural frequencies are
too spread as to be part of the coherent pack. Finally, after further
increasing the value of $\lambda$, more and more nodes get entrained
around the mean phase and the system settles in a completely
synchronized state where $r\approx1$ (not shown).

A comparison between the results for different values of $\mu$ (and
thus different $\langle c \rangle$ and $\langle L \rangle$ values)
indicate several interesting features of the synchronization
process. First, it is remarkably that when the clustering coefficient
increases, the system reaches {\em complete synchronization} at higher
values of the coupling. This result agrees with the results reported
in \cite{mcgraw}, where a different network model able to generate
topologies with a tunable clustering coefficient was implemented. 

At this point, one may ask whether the effects are only due to the
influence of $\langle c \rangle$ or to the increase of the average
path length \cite{note2} (note that the model implemented in
\cite{mcgraw} does not explore this possibility). Unfortunately, the
two factors are generally linked together so they can not be
considered separately. However, as stated previously, a closer look at
Fig.\ \ref{fig1} reveals that there is a region of the parameter $\mu$
where the clustering coefficient grows while the average shortest path
length remains almost constant. This corresponds to the interval $0.4
\le \mu \le 1.0$ approximately. Going back to Fig.\ \ref{fig2}, the
behavior of $r$ in this interval of $\mu$ reveals that synchronization
is almost unaffected. In fact, the $r(\mu)$ curves lie slightly above
that corresponding to the BA limit. Therefore, though the above
comparison is not conclusive, it seems that the delayed transition to
complete synchronization is mainly due to the effect of the increase
in $\langle L \rangle$ at smaller values of $\mu$ rather than to the
increase in $\langle c \rangle$. This conclusion is further supported
by a direct comparison of the results in Fig.\ \ref{fig2} with those
reported in \cite{mcgraw}, where the authors explored a region with
higher values of $\langle c \rangle$ (up to 0.7) and the profile of
$r(\lambda)$ is almost the same as ours.

\begin{figure*}[!htb]
\begin{center}
\epsfig{file=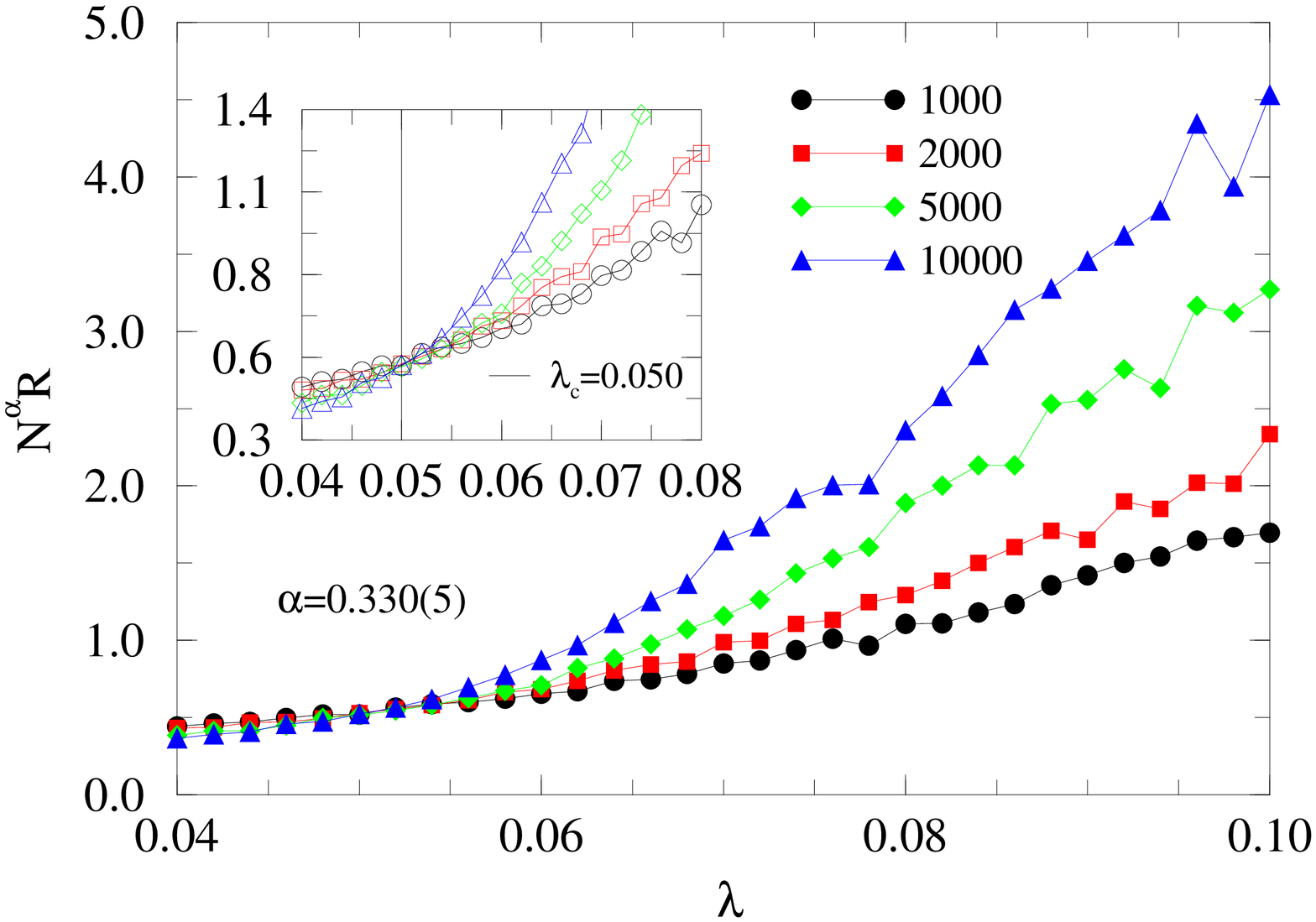,width=2.3in,angle=-90}
\epsfig{file=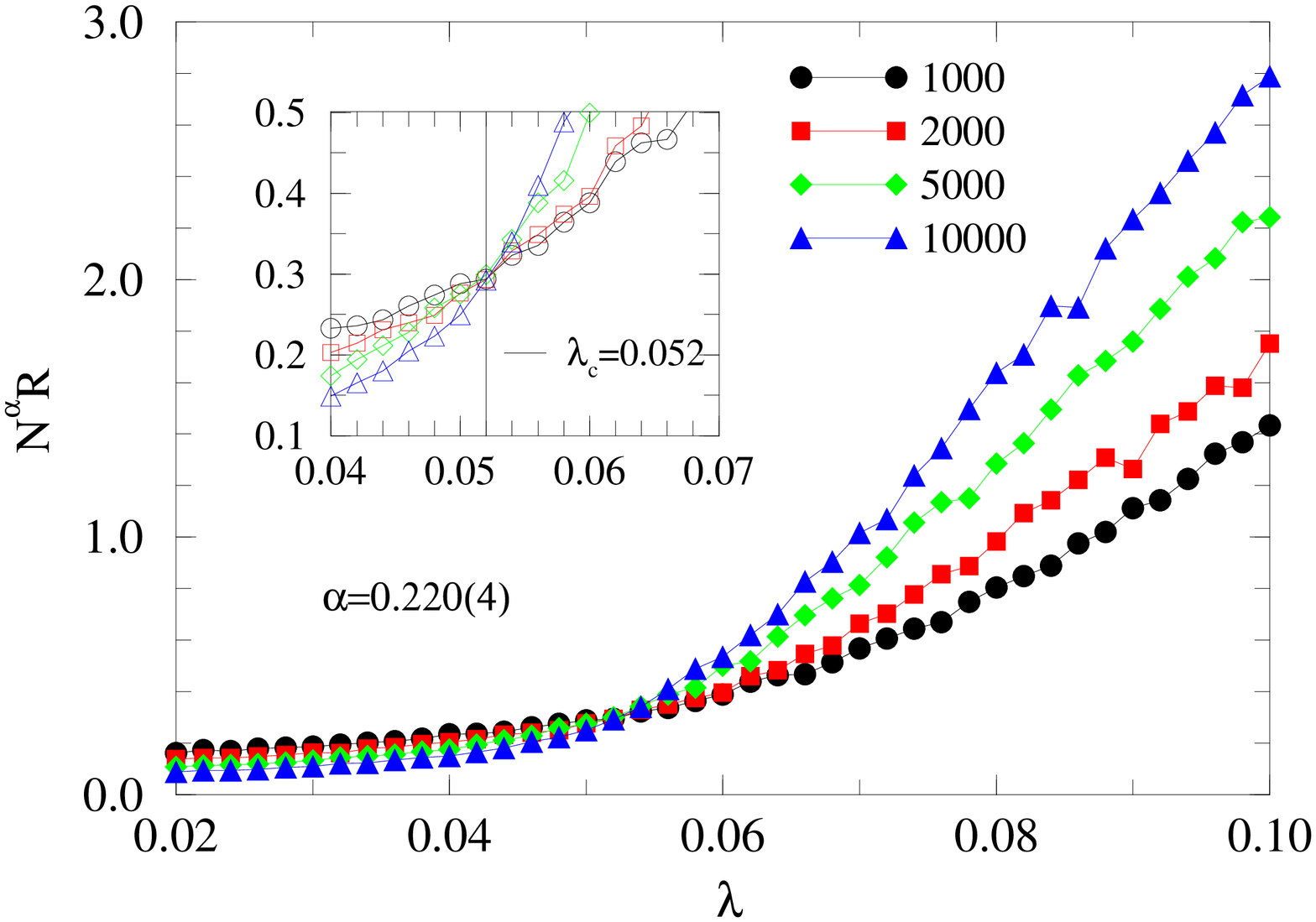,width=2.3in,angle=-90}
\epsfig{file=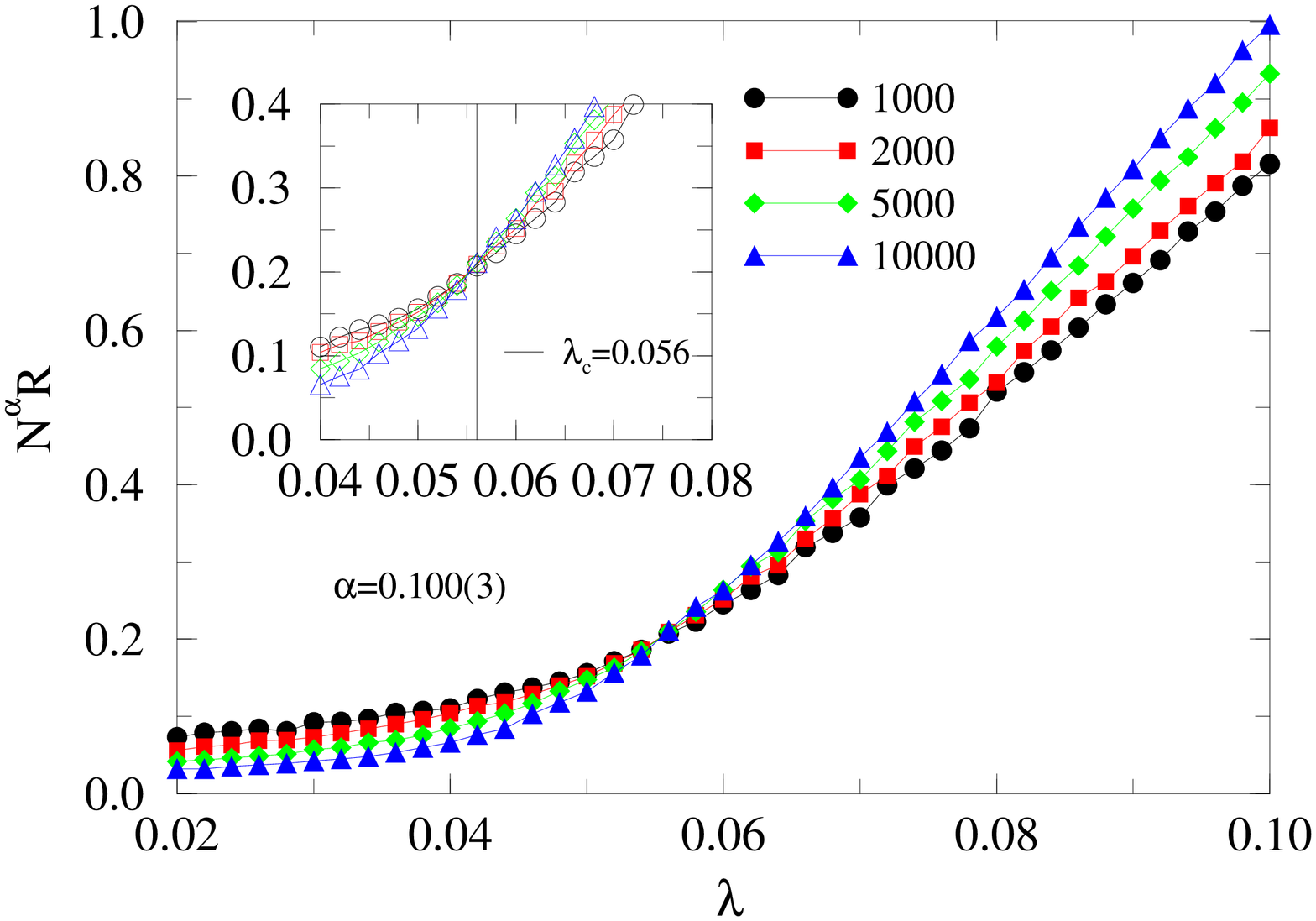,width=2.3in,angle=-90}
\epsfig{file=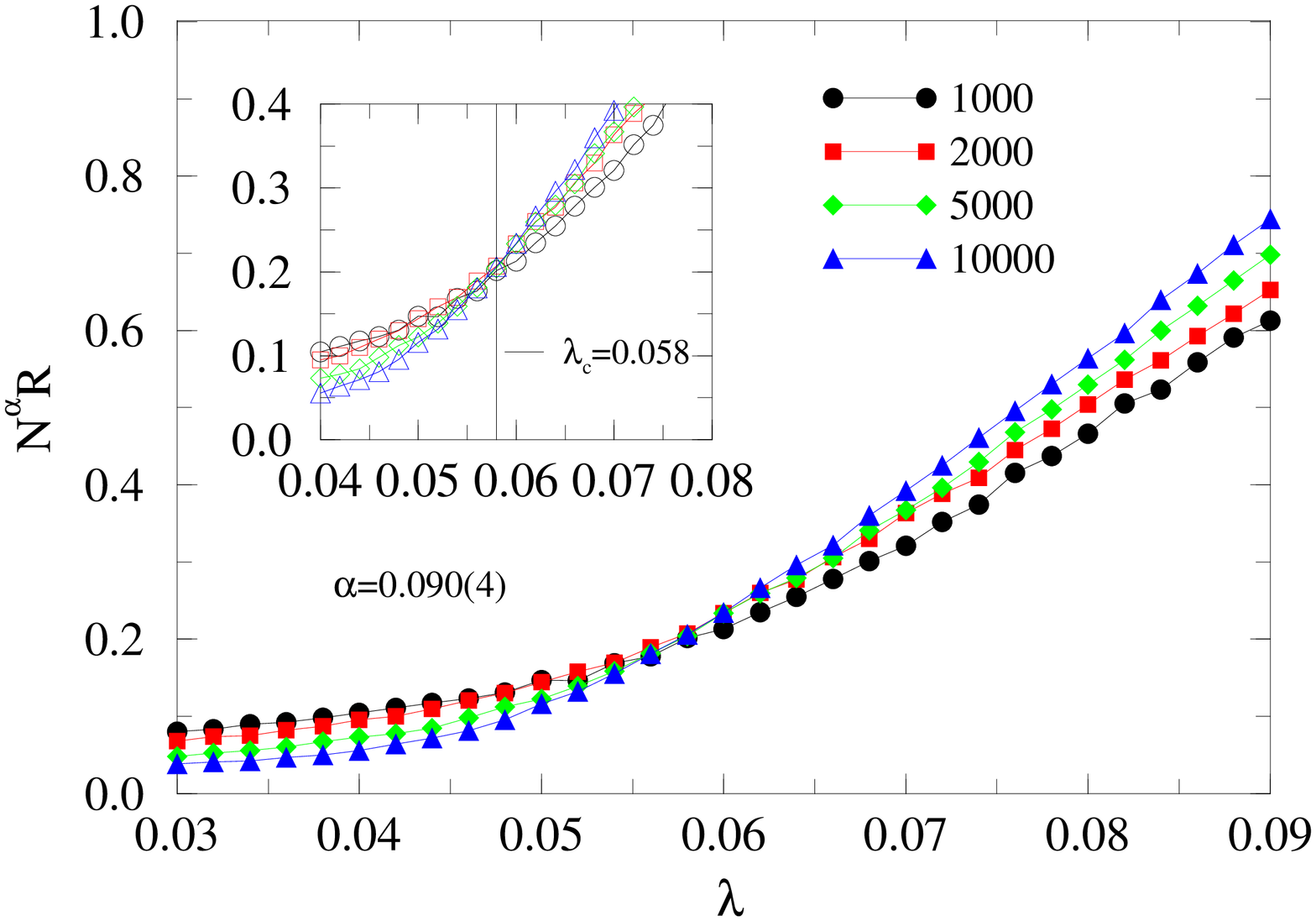,width=2.3in,angle=-90}
\end{center}
\caption{(color online) Finite size scaling analysis for several
  values of $\mu$. From top to bottom and from left to right the
  values of $\mu$ are: 0.05, 0.15, 0.50 and 0.60. In each panel, it is
  represented the rescaled order parameter against the control
  parameter $\lambda$. The insets are a zoom to the regions around the
  critical points $\lambda_c(\mu)$. The data are averaged over at
  least 100 realizations for each value of $\lambda$. The sizes of the
  networks, the critical points $\lambda_c(\mu)$ at which the onset of
  synchronization takes place, as well as the values of the critical
  exponents $\alpha$ are those indicated in the plots. See the main
  text for more details.}
\label{fig3}
\end{figure*}

The second region of interest is the onset of synchronization. From
Fig.\ \ref{fig2}, it is difficult to elucidate how the critical point
for the BA limit compares with those at values of $\mu <1$. At first
glance, it seems that $\lambda_c(\mu)$ shifts rightward as the
parameter $\mu$ is decreased below 1. However, a more detailed
analysis shows that it is indeed the contrary. To this end, we have
performed a finite size scaling analysis that allows to determine the
critical points $\lambda_c(\mu)$. We assume a scaling relation of the
form
\begin{equation}
r=N^{-\alpha}f(N^{\beta}(\lambda-\lambda_c)),
\end{equation}
where $f(x)$ is a universal scaling function bounded as $x \rightarrow \pm
\infty$ and $\alpha$ and $\beta$ are critical exponents to be
determined. The estimation of $\lambda_c$ can then be done by
plotting $N^{\alpha}r$ as a function of $\lambda$ and tuning
$\alpha$ for several system sizes $N$ until the curves cross at a
single point, the critical one. 

The results of the FSS analysis are shown in Fig.\ \ref{fig3} for
different values of $\mu$ (from top to bottom and from left to right
$\mu=0.05,0.15,0.50,0.60$). The insets show a blow-up around the
critical points $\lambda_c(\mu)$. Although the differences in the
critical points at different values of $\mu$ are small, they are
certainly distinguishable. In fact, the higher the value of $\mu$, the
higher the critical point. That is, when the clustering coefficient
and the average path length grow with respect to the BA network, the
onset of synchronization is anticipated. Moreover, taking into account
that the increase in $\langle L \rangle$ is likely to inhibit
synchronization, one may hypothesize that the effects of the
clustering coefficient prevail in this region of the parameter
$\lambda$. To check this hypothesis, we have also included in Fig.\
\ref{fig3} the analysis performed for $\mu=0.50$ and $\mu=0.60$. As
pointed out before, for these values, the differences can only arise
from the variations of the clustering coefficient as the average path
length remains constant in this region of the parameter $\mu$. The
critical points, although very close to each other, are clearly
different. Therefore, the main contribution to the onset of
synchronization at low values of $\lambda$ comes from the raising of
the clustering coefficient.

\section{Discussions and Conclusions}

Rounding off, our results point to a nontrivial dependence between the
clustering coefficient and the average path length, and the
synchronization patters of phase oscillators. Separately, the onset of
synchronization seems to be mainly determined by $\langle c \rangle$,
promoting synchronization at low values of the coupling strength with
respect to networks not showing high levels of structural
clustering. On the other hand, when the coupling is increased beyond
the critical point, the effect of $\langle L \rangle$ dominates and
the phase diagram is smoothed out (a sort of stretching), delaying the
appearance of the fully synchronized state. These results confirm and
complement those anticipated in \cite{mcgraw} and show that general
statements about synchronizability using the MSF are
misleading. Whether or not a system is more or less synchronizable
than others showing distinct structural properties is relative to the
region of the phase diagram in which the system operates
\cite{jma01,jma02}.

In summary, we have shown that synchronizability of complex networks
is dependent on the effective coupling $\lambda$ among oscillators,
and on the properties of the underlying network. For small values of
$\lambda$, the incoherent solution $r=0$ first destabilizes as the
clustering coefficient is higher, while the coherent solution $r=1$ is
promoted when both the structural clustering and the average path
length are small. Finally, we point out that our results are also
consistent if a different local order parameter is considered
\cite{jma02}. Moreover, though these results have been obtained for
phase oscillators, we think that they should hold for other nonlinear
dynamical systems as well. It would be interesting to check this later
hypothesis in future works.

\begin{acknowledgments}
  We thank A. Arenas for helpful comments and discussions on the
  subject. J.G.G. and Y.M. are supported by MEC through a FPU grant
  and the Ram\'{o}n y Cajal Program, respectively. This work has been
  partially supported by the Spanish DGICYT Projects
  FIS2004-05073-C04-01 and FIS2005-00337.
\end{acknowledgments}


\begin{thebibliography}{00}
\bibitem{note1} We refer to Refs \cite{jma01,jma02} for a discussion
  on the proper choice of the coupling strength describing the
  interaction between the oscillators.
\bibitem{note2} Note that although the main differences in the network
model are observed for $\langle c \rangle$ and $\langle L \rangle$,
other properties may slightly vary as well.
\end{thebibliography}
\end{document}